\documentclass[twocolumn,showpacs,preprintnumbers,amsmath,amssymb,nofootinbib,aps,pra,10pt]{revtex4-1}
\usepackage{graphicx}%
\usepackage{dcolumn}%
\usepackage{bm}%
\usepackage{color}
\usepackage[pdftex,
    pdftitle={Elliptic fibrations for SU(5) x U(1) x U(1) F-theory vacua},
    pdfauthor={Jan Borchmann, Christoph Mayrhofer, Eran Palti, Timo Weigand},
    pdfsubject={F-theory compactifications},
    bookmarksopen, bookmarksnumbered, bookmarksopenlevel=2]{hyperref}

\newcommand{\be}{\begin{equation}}
\newcommand{\ee}{\end{equation}}
\newcommand{\bea}{\begin{eqnarray}}
\newcommand{\eea}{\end{eqnarray}}

\newcommand{\T}[1]{\textmd{#1}}

\newcommand{\tw}{\text{w}}
\newcommand{\tv}{\text{v}}
\newcommand{\tu}{\text{u}}

\newcommand{\executeiffilenewer}[3]{%
 \ifnum\pdfstrcmp{\pdffilemoddate{#1}}%
 {\pdffilemoddate{#2}}>0%
 {\immediate\write18{#3}}\fi%
}

\begin{document}

\title{Elliptic fibrations for $SU(5) \times U(1) \times U(1)$ F-theory vacua
}%
\author{
             Jan Borchmann$^{1}$, Christoph Mayrhofer$^{1}$,
             Eran Palti$^{2}$ and 
             Timo Weigand$^{1}$}
\affiliation{%
$^{1}$ Institut f\"ur Theoretische Physik, Ruprecht-Karls-Universit\"at Heidelberg,
             Germany
\\
$^{2}$ Centre de Physique Theorique, Ecole Polytechnique,
               CNRS, %
               Palaiseau, France
}

\begin{abstract}
Elliptic Calabi-Yau fibrations with Mordell-Weil group of rank two are constructed. Such geometries are the basis for F-theory compactifications with two abelian gauge groups in addition to non-abelian gauge symmetry.
We present the elliptic fibre both as  a ${\rm Bl}^2 \mathbb P^2[3]$-fibration and in the birationally equivalent Weierstra\ss{} form. 
The spectrum of charged singlets and their Yukawa interactions are worked out in generality. This framework can be combined with the toric construction of tops to implement additional non-abelian gauge groups.
We utilise the classification of tops to construct $SU(5)\times U(1) \times U(1)$ gauge symmetries systematically and study the resulting geometries,  presenting the defining equations, the matter curves and their charges,  the Yukawa couplings and explaining the process in detail for an example. Brane recombination relates these geometries to a ${\rm Bl}^1 \mathbb P^2[3]$-fibration with a corresponding class of $SU(5) \times U(1)$ models.
We also present the 
$SU(5)$ tops based on 
the elliptic fibre  ${\rm Bl}^1 \mathbb P_{[1,1,2]}[4]$, corresponding to another class of $SU(5) \times U(1)$ models. 
\end{abstract}
\maketitle

\section{Introduction}

Recently a lot of progress has been made in the construction of elliptically fibered Calabi-Yau 4-folds leading to four-dimensional F-theory compactifications with abelian gauge groups \cite{Grimm:2010ez,Braun:2011zm,Krause:2011xj,Grimm:2011fx,Morrison:2012ei,Mayrhofer:2012zy,Braun:2013yti}. 
Such constructions are motivated in part by the manifold applications of abelian gauge symmetry in string model building. These include the prominent role of $U(1)$ selection rules in
phenomenology and their relevance in particular for Grand Unified Theory (GUT) model building, where $U(1)$ selection rules can be responsible for proton stability, prevent too large $\mu$-terms or induce realistic flavour structure \cite{Marsano:2009gv,Dudas:2010zb,Dolan:2011iu,Palti:2012dd,Mayrhofer:2013ara,Weigand:2010wm}. Furthermore, the construction of $U(1)$ symmetries endows one with a large class of gauge fluxes required to generate a chiral matter spectrum \cite{Braun:2011zm,Krause:2011xj,Grimm:2011fx,Krause:2012yh}.

F-theory compactifications with abelian symmetries are based on elliptically fibred Calabi-Yau 4-folds with Mordell-Weil group of rank one or bigger \cite{Morrison:1996na}.
The rank of the Mordell-Weil group gives the number of independent rational sections  of the fibration. While every elliptic 4-fold suitable for F-theory necessarily exhibits a universal holomorphic section that identifies the base $B_3$ of the fibration as the physical compactification space, extra rational sections are related to certain elements of $H^{1,1}(\hat Y_4)$, other than the class dual to $B_3$, which do not lie in $H^{1,1}(B_3)$.  Such 2-forms $\tw_i$ give rise to a $U(1)$ gauge potential upon expanding the M-theory 3-form $C_3$ as $C_3 = A_i \wedge \tw_i$ \cite{Morrison:1996na}.

In this letter,  we report on the %
construction of elliptically fibred Calabi-Yau 4-folds $\hat Y_4$ with Mordell-Weil group of rank 2. 
To construct fibrations with two independent extra sections, we consider an elliptic fibre described as %
a slightly restricted cubic in $\mathbb P^2$, cf.\ eq.\ \eqref{eq:hyper1} and FIG.~\ref{fig:polygon5}. 
This fibration gives rise to $U(1) \times U(1)$ gauge symmetry with charged singlet states, whose structure and Yukawa interactions we present.

In addition to this universal $U(1) \times U(1)$ charged singlet sector, extra non-abelian gauge symmetries along divisors can be engineered.
Based on the toric classification \cite{Bouchard:2003bu} of tops \cite{Candelas:1996su}, we have explicitly worked out \cite{new} the tops
leading to an $SU(5)$ 
singularity over a divisor on $B_3$, corresponding in total to F-theory compactifications with $SU(5) \times U(1) \times U(1)$ gauge group. We present the fully resolved 4-folds and discuss the matter spectrum and the Yukawas.

 In this letter we also provide the tops leading to $SU(5)$ $\times U(1)$-fibrations based on the elliptic fibre $\mathbb {\rm Bl^1}\mathbb P_{[1,1,2]}[4]$ \cite{Morrison:2012ei}, extending our previous studies
\cite{Mayrhofer:2012zy}.

It is interesting to note that our $SU(5) \times U(1) \times U(1)$-fibrations lend themselves also to studying models with a single $U(1)$ factor upon Higgsing a linear combination of the two $U(1)$s.
This brane recombination process leads to a ${\rm Bl}^1 \mathbb P^2[3]$-fibration, and the charges and GUT matter curves of the associated class of $SU(5) \times U(1)$ fibrations include e.g.\ the model presented recently in \cite{Braun:2013yti}.

Our analysis does not specify the base space of the fibration and is thus applicable very generally.
While we reserve a detailed description and full display of our results to the companion paper \cite{new}, here we outline the main features of our approach and exemplify it with one specific fibration of the above type. 

\section{\texorpdfstring{${\rm Bl}^2 \mathbb P^2[3]$}{Bl2 P2[3]}-fibrations}

The starting point of our construction is the representation of an elliptic curve as the cubic hypersurface
\begin{equation}
\begin{split} \label{eq:hyper1}
0&=  \tv\, \tw (c_1\,  \tw\, + c_2\, \tv )  + \tu\, (b_0\, \tv^2\, + b_1\, \tv\, \tw + b_2\, \tw^2) +  \\
    &     +  \tu^2 (d_0\, \tv  + d_1\, \tw + d_2\, \tu)
\end{split}
\end{equation}
with $[\tu:\tv:\tw]$ homogeneous coordinates of  $\mathbb P^2$. If we promote   $d_i$, $c_i$ and $b_i$ to sections of  suitable line bundles on a 3-dimensional base $B_3$, this defines an elliptically  fibred Calabi-Yau 4-fold $Y_4$. %
The hypersurface \eqref{eq:hyper1} is a non-generic cubic within $\mathbb P^2$ to the extent that the coefficient of $\tw^3$ and $\tv^3$ are set to zero. As a result, the  elliptic fibre \eqref{eq:hyper1} contains the points 
\begin{equation}
\begin{split}
\label{eq:sections_2u1s_cubic}
 \T{Sec}_0:\qquad [\tu\,:\,\tv\,:\,\tw] &= [0\,:\,0\,:\,\tw],\\
 \T{Sec}_1:\qquad [\tu\,:\,\tv\,:\,\tw] &= [0\,:\,\tv\,:\,0],\\
 \T{Sec}_2:\qquad [\tu\,:\,\tv\,:\,\tw] &= [0\,:\,-c_1\,:\,c_2].
\end{split}
\end{equation}
It is instructive to bring \eqref{eq:hyper1}, by means of a birational map \cite{Nagell,new}, into Weierstra\ss{} form $y^2 = x^3 + f x z^4 + g z^6$,  
where the sections $f$ and $g$ are given by
\begin{equation}
 f=-\tfrac13\T d^2 + \T c\, \quad\T{and}\quad g=-  f\,\left(\tfrac13\T d\right) -\left(\tfrac{1}{3}\T d\right)^3  + \T e \nonumber
\end{equation}
with
\begin{equation}
 \begin{split}
  \T d &=b_1^2 + 8\,b_0\,b_2 - 4\,c_1\,d_0 - 4\,c_2\,d_1,\\
  \T c &= -8  \big({ b_0 \left( b_1  c_1  d_1-b_1^2 b_2+2  b_2  c_1  d_0+2  b_2  c_2
    d_1-2  c_1^2  d_2\right)}  \\
    &+{\left( c_2 ( b_1  b_2  d_0+ b_1
    c_1  d_2-2  b_2  c_2  d_2-2  c_1  d_0
    d_1)-2  b_0^2  b_2^2\right)}  \big),\\
  \T e &= 16  ( b_0  b_1  b_2- b_0  c_1  d_1- b_2  c_2
    d_0+ c_1  c_2  d_2)^2  . \nonumber
 \end{split}
\end{equation}
Under this map the point $\T{Sec}_0$ maps to the ``zero point'' $[x:y:z] =[\lambda^2:\lambda^3:0]$ of the Weierstra\ss{} model.  
Therefore $\T{Sec}_0$ is related to the universal section, while $\T{Sec}_1$ and  $\T{Sec}_2$ are related to the extra rational sections responsible for the two $U(1)$ gauge groups. 
To make this relation explicit we first need to resolve the conifold singularities exhibited by the fibration for generic $d_i$, $c_i$ and $b_i$.

In fact, from the Weierstra\ss{} representation of \eqref{eq:hyper1}  we find \cite{new} that the elliptic fibre develops $SU(2)$ singularities over  the  codimension-two loci on $B_3$ given by
\begin{equation}\label{eq:conifold_locus_1a_explicit} 
\begin{split}
d_0\, c_2^2 & = b_0 \,b_1 \, c_2 -b_0^2\, c_1 \,,\\
d_1\, b_0\, c_2 & = b_0^2\, b_2 + c_2^2\, d_2
\end{split}
\end{equation}
and
\begin{equation}\label{eq:conifold_locus_1b_explicit} 
\begin{split}
d_1\, c_1^2 & = b_1\, b_2\, c_1 - b_2^2\, c_2\,,\\
d_0\, b_2\, c_1 & = b_0\, b_2^2 + c_1^2\, d_2\,
\end{split}
\end{equation}
as well as
\begin{equation}\label{eq:conifold_locus_2_explicit} 
\begin{split}
 c_1^3\,\big(d_0\, c_2^2  - &  b_0\, b_1\, c_2  +b_0^2\, c_1 \big)=\\
& \qquad =  c_2^3\, \big( d_1\,c_1^2 -b_1\, b_2\, c_1 + b_2^2\, c_2\big)\,,\\
d_2\, c_1^4\, c_2^2  =&  \left(c_2\, (b_1\, c_1 - b_2\, c_2)-b_0\, c_1^2\right) \big(b_0\, b_2\, c_1^2 +  \\
&\quad +c_2 \big( d_1\,c_1^2 -b_1\, b_2\, c_1 + b_2^2\, c_2\big)\big)\,.
\end{split}
\end{equation}
These in turn have the following solutions:
\begin{enumerate}
\item $C_{\mathbf 1^{(1)}}$: $b_0=c_2=0$; \label{b0c2}
\item $C_{\mathbf 1^{(2)}}$: \eqref{eq:conifold_locus_1a_explicit} with $(b_0, c_2) \neq (0,0)$;
\item $C_{\mathbf 1^{(3)}}$: $b_2=c_1=0$; \label{b2c1}
\item $C_{\mathbf 1^{(4)}}$: \eqref{eq:conifold_locus_1b_explicit} with $(b_2, c_1) \neq (0,0)$;
\item $C_{\mathbf 1^{(5)}}$:   $c_1=c_2=0$;
\item $C_{\mathbf 1^{(6)}}$:  \eqref{eq:conifold_locus_2_explicit} with $(c_1,c_2) \neq (0,0)$, %
$(b_0, c_2) \neq (0,0)$ and $(b_2, c_1) \neq (0,0)$. %
\end{enumerate} 

\begin{figure}[h]
\begin{flushright}
\def\svgwidth{0.4\textwidth}
 \executeiffilenewer{polygon5.svg}{polygon5.pdf}%
 {inkscape -z -D --file=polygon5.svg %
  --export-pdf=polygon5.pdf --export-latex}%
   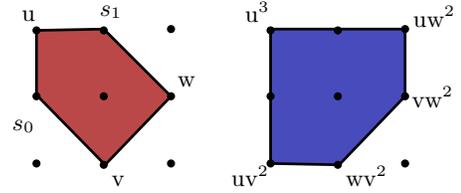%

      \caption{The toric polygon to $\textmd{Bl}^{2}\mathbb{P}^{2}$ and its dual. On the dual one we only indicated the monomials of the vertices and omitted powers of $s_0$ and $s_1$. }\label{fig:polygon5}
\end{flushright}
\end{figure}

The singularities are resolved by two blow-ups:
The fibre over the curve $C_{\mathbf 1^{(1)}}$  is singular in the point $\T{Sec}_1$, which is remedied by introducing the blow-up coordinate $s_1$ via
\bea
\tu \rightarrow  s_1\,\tu, \qquad \tw \rightarrow  s_1\,\tw.
\eea
Likewise, the singularity at $\T{Sec}_0$ in the fibre over $C_{\mathbf 1^{(3)}}$ is resolved via
\bea
\tu \rightarrow  s_0\,\tu, \qquad \tv \rightarrow  s_0\,\tv.
\eea
The proper transform of \eqref{eq:hyper1} after these two blow-ups reads
\begin{eqnarray} \label{eq:hyper1-res}
&& \tv\, \tw (c_1\,  \tw\, s_1 + c_2\, \tv\,s_0)  + \tu\, (b_0\, \tv^2\,s_0^2 + b_1\, \tv\, \tw\,s_0\,s_1 + \qquad\\
&& b_2\, \tw^2\,s_1^2) + 
   \tu^2 (d_0\, \tv \,s_0^2\,s_1 + d_1\, \tw \,s_0\,s_1^2 + d_2\, \tu\,s_0^2\,s_1^2)=0\nonumber
\end{eqnarray}
and indeed describes a smooth manifold $\hat Y_4$ for generic base sections. This can be checked by exploiting the enlarged Stanley-Reisner ideal %
\begin{equation}
 \{ \T w \, s_0, \T w \, \T u,  \T v \, s_1,  s_0 \, s_1,\T v \, \T u\}. \label{SR1}
\end{equation}
The resolved fibre ambient space is shown in FIG.~\ref{fig:polygon5}.
The section $s_0=0$ can be viewed as the universal section of $\hat Y_4$, while $s_1=0$ and $\tu=0$---the section corresponding to $\T{Sec}_2$---are the generators of the Mordell-Weil group. 
We will denote these sections as 
\bea
S_0: \, s_0 = 0, \qquad  S_1: \, s_1=0, \qquad S_2: \, \tu=0\,.
\eea

\begin{figure}[h]
\begin{center}
\def\svgwidth{0.4\textwidth}
 \executeiffilenewer{singlets_polygon5.svg}{singlets_polygon5.pdf}%
 {inkscape -z -D --file=singlets_polygon5.svg %
  --export-pdf=singlets_polygon5.pdf --export-latex}%
   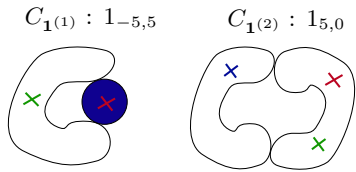%

      \caption{The fibre structure over the singlet curves $C_{\mathbf 1^{(1)}}$ and $C_{\mathbf 1^{(2)}}$. Green corresponds to the zero section $S_0$, blue to $S_1$ and red to $S_2$.}\label{fig:singlets_polygon5}
\end{center}
\end{figure}

The fibre over each of the six curves $C_{\mathbf 1^{(i)}}$ splits into two rational curves $\mathbb P^1_A$ and $\mathbb P^1_B$ which intersect in two points, corresponding to the affine Dynkin diagram of $SU(2)$.
Consider for example the curve $C_{\mathbf 1^{(1)}}$. The resolved hypersurface \eqref{eq:hyper1-res} at  $b_0=c_2=0$ factorises as $s_1 (\ldots) =0$. The section $S_1$ therefore wraps, say, $\mathbb P^1_A$
of the fibre, while $(\ldots)=0$ describes the second $\mathbb P^1_B$. Since $\{s_0, s_1\}$ are in the Stanley-Reisner ideal, $S_0$ intersects only $\mathbb P^1_B$---in one point. Furthermore $S_2$ intersects $\mathbb P^1_A$ in one point. This follows by counting common points of the various equations.  In a similar way the topology over the remaining 5 curves can be understood. This behaviour is depicted for  $C_{\mathbf 1^{(1)}}$ and $C_{\mathbf 1^{(2)}}$ in FIG.~\ref{fig:singlets_polygon5}.

M2-branes wrapping either of the $\mathbb P^1$s in the fibre over $C_{\mathbf 1^{(i)}}$ give rise to singlet states $\mathbf 1^{(i)}$ charged under the two $U(1)$s of the model. To compute their charges we first need to 
construct the 2-forms $\tw_i \in H^{1,1}(\hat Y_4)$  responsible for the two $U(1)$ gauge potentials via the Shioda map. %
By F/M-theory duality these must satisfy the transversality condition 
\begin{equation} 
\int_{\hat Y_{4}}\text{w}_{i}\wedge \omega_{a} %
=0,\quad \int_{\hat Y_{4}}\text{w}_{i}\wedge S_0\wedge  %
\omega_{b}=0 \nonumber
\end{equation}
for all closed 6-forms $\omega_a$ and 4-forms $\omega_b$ of the base of the fibration pulled back to $\hat Y_4$, where for brevity we omit the pullback symbol. 
We find \cite{new} that the conveniently normalised\footnote{The overall factor 5 appears because in section \ref{sec:su5-model} we are considering $SU(5)$ models.}
\begin{align}
\begin{split} 
\text{w}_{1} = 5( S_{1} - S_{0} - \mathcal{\bar K}), \quad \text{w}_{2} = 5 (S_2 - S_{0} - \mathcal{\bar K} - [c_{1}]) \nonumber
\end{split}
\end{align}
satisfy these constraints, where ${\bar {\cal K}}$ is the anti-canonical class of $B_3$ and $[c_1]$ the dual 2-form associated with the base section $c_1$.  This follows from the fact that $S_i$ are sections together with the intersection numbers
\bea
&& \int_{\hat Y_{4}}S_2 \wedge S_{0}\wedge \omega_b = \int_{\hat Y_{4}}[c_{1}]\wedge S_{0}\wedge \omega_b, \nonumber \\
&& \int_{\hat Y_{4}}S_{1}\wedge S_{0}\wedge \omega_b = 0.
\eea

The charges of the singlets follow by integrating $\tw_i$ over either of the two $\mathbb P^1$s of the fibres. By means of the specific intersections of the sections $S_i$ with the $\mathbb P^1$s, we can evaluate these integrals. Note that the classes $\bar{\cal K}$ and $[c_1]$ do not contribute. 
The result is
\begin{equation}\label{Singlet-charges}
\begin{aligned}
C_{\mathbf 1^{(1)}}: \mathbf 1_{-5,5},\,\,\,\,\,  && C_{\mathbf 1^{(2)}}: \mathbf 1_{5,0}, \quad && C_{\mathbf 1^{(3)}}: \mathbf 1_{5,10}, \\
 C_{\mathbf 1^{(4)}}: \mathbf 1_{-5,-5},\, && C_{\mathbf 1^{(5)}}: \mathbf 1_{0,-10}, && C_{\mathbf 1^{(6)}}: \mathbf 1_{0,5} ,\,\,
\end{aligned}
\end{equation}
plus the conjugate states in each case.

\begin{figure}[h]
\begin{center}
\def\svgwidth{0.35\textwidth}
 \executeiffilenewer{singlets_Yukawas_polygon5.svg}{singlets_Yukawas_polygon5.pdf}%
 {inkscape -z -D --file=singlets_Yukawas_polygon5.svg %
  --export-pdf=singlets_Yukawas_polygon5.pdf --export-latex}%
   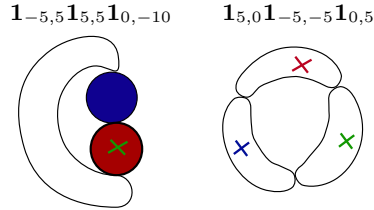%

      \caption{The fibre structure over the Yukawa points $\mathbf 1_{-5,0}$ $\mathbf 1_{5,10}   \mathbf 1_{0,-10}$ and $\mathbf 1_{5,0}   \mathbf 1_{-5,-5}  \mathbf 1_{0,5}$. Green corresponds to the zero section $S_0$, blue to $S_1$ and red to $S_2$.}\label{fig:singlets_Yukawas_polygon5}
\end{center}
\end{figure}

As a new feature not present in previous F-theory models with only one $U(1)$ gauge group factor, 
the charged singlets possess Yukawa couplings with each other. This happens whenever two or three singlet curves intersect in a point.
For generic base sections there are three types of such codimension-three loci on $B_3$,
\begin{equation}
\begin{split}
 C_{\mathbf 1^{(1)}} \cap C_{\mathbf 1^{(4)}}  \cap C_{\mathbf 1^{(5)}} & = \{ b_0 = c_2 = c_1 = 0 \}, \\
 C_{\mathbf 1^{(2)}}  \cap C_{\mathbf 1^{(3)}} \cap C_{\mathbf 1^{(5)}} & = \{ b_2 = c_1 = c_2 = 0 \}, \\ 
 C_{\mathbf 1^{(2)}} \cap C_{\mathbf 1^{(4)}}  \cap C_{\mathbf 1^{(6)}} & = \{ \ldots \},
\end{split}
\end{equation}  
where the last expression is a bit more lengthy but nothing else than the first eq.\ of \eqref{eq:conifold_locus_1a_explicit} and \eqref{eq:conifold_locus_1b_explicit} and the second eq.\ of \eqref{eq:conifold_locus_2_explicit}.
Over $C_{\mathbf 1^{(2)}}  \cap C_{\mathbf 1^{(3)}} \cap C_{\mathbf 1^{(5)}}$ and $C_{\mathbf 1^{(1)}} \cap C_{\mathbf 1^{(4)}}  \cap C_{\mathbf 1^{(5)}} $ the hypersurface factorises into $\tu\, s_0\, (\ldots)$  and into $\tu\, s_1\, (\ldots)$, respectively. 
Likewise, over $C_{\mathbf 1^{(2)}} \cap C_{\mathbf 1^{(4)}}  \cap C_{\mathbf 1^{(6)}}$ the hypersurface factorises also into three components. All of these three components are  $\mathbb P^1$s.
With the help of the Stanley-Reisner ideal one confirms that these intersect like the nodes of the affine Dynkin diagram of $SU(3)$, cf.\ FIG.~\ref{fig:singlets_Yukawas_polygon5}. %
This leads to the following structure of singlet interactions:
\begin{center}
\begin{tabular}{c|c}
Point & Yukawa coupling \\
\hline
$\{ b_0 = c_2 = c_1 = 0 \}$                   &  $\mathbf 1_{-5,5}  \mathbf 1_{5,5}    \mathbf 1_{0,-10}$\\
$\{ b_2 = c_1 = c_2 = 0 \}$                   &  $\mathbf 1_{-5,0}  \mathbf 1_{5,10}   \mathbf 1_{0,-10}$\\
$C_{\mathbf 1^{(2)}}\cap C_{\mathbf 1^{(4)}}\cap C_{\mathbf 1^{(6)}}$ &  $\mathbf 1_{5,0}   \mathbf 1_{-5,-5}  \mathbf 1_{0,5}$  
\end{tabular}
\end{center}

\section{ An \texorpdfstring{$SU(5)\times U(1)^2$}{SU(5)xU(1)2} model}
\label{sec:su5-model}

The fibration \eqref{eq:hyper1-res}  fits naturally in the class of toric elliptic fibrations considered in \cite{Bouchard:2003bu}. %
The toric algorithm of tops %
allows for the systematic construction of non-abelian singularities over a divisor $w=0$ in $B_3$ %
which are torically possible. This amounts to the possible ways how the otherwise generic base sections $b_i$, $c_i$ and $d_i$ can factor out suitable overall powers of $w$ such that the birationally equivalent Weierstra\ss{} model exhibits the desired singularity. These vanishing degrees are encoded in the toric data as so-called tops \cite{Candelas:1996su}. In fact \cite{Bouchard:2003bu} classifies the tops for all  $T^2$-fibrations where the torus fibre is given by a reflexive polygon, i.e.\ where the fibre can be represented as a hypersurface in a 2-dimensional toric variety. The polygon in FIG.~\ref{fig:polygon5} is listed there as number five out of sixteen possibilities.  For gauge group $SU(5)$ one finds nine $SU(5)$ tops for the ${\rm Bl}^2 \mathbb P^2[3]$-fibration, each corresponding to a particular assignment of vanishing degrees of the base sections.

As one out of the nine possible enhancement patterns, we now consider 
 the following described by the hypersurface  %
\begin{equation}
\begin{split}\label{hyperSU51}
0& =  b_{0,2} w^2 s_0^2 \T v^2  \T u +c_{2,1} w s_0 \T w \T v^2+ d_{0,2} w^2 \T v s_0^2 s_1 \T u^{2} + \\
& +  b_1 s_0 s_1 \T w \T v \T u + 
c_1 \T w^{2} \T v s_1 +  d_{2,2} w^2  s_0^2  s_1^2 \T u^{3} +  \\
&+  d_1    s_0 s_1^2 \T w \T u^{2}+b_2   s_1^{2} \T w^2 \T u.
\end{split}
\end{equation}
The defining data of the remaining possibilities are presented in appendix \ref{app-SU5tops1} and described in more detail in \cite{new}.
From the associated discriminant $\Delta = w^{5}(P + \mathcal{O}(w))$ (computed most easily in the birationally equivalent Weierstra\ss{} model) with
\begin{equation}
 \begin{split}
 P =& \frac{1}{16}\, b_1^{4}  b_{0,2}\,  c_{2,1} \, c_1\, ( b_1 b_2 -  d_1 c_1)\\
&\qquad\qquad\times \left( d_{2,2} b_1^{2} +  d_1( b_{0,2} d_1 -  d_{0,2} b_1)\right)  
 \end{split}
\end{equation}
we confirm the gauge group $SU(5)$ along $w=0$, henceforth called GUT divisor, and anticipate %
the existence of a $\mathbf{10}$-matter curve
\begin{equation}
C_{\mathbf{10}} = \{b_1 = 0\} \cap \{w =0\}
\end{equation}
as well as five ${\mathbf 5}$-matter curves\footnote{To distinguish between the $\mathbf{10}$- and  $\mathbf{5}$-curve we would also need the next order in $w$ of the discriminant, which we do not display here for brevity. Note that in the following we analyse the enhancement loci in the resolved Calabi-Yau, from which we observe the difference as well.}
\bea
&& C_{ \mathbf 5^{(1)}} = \{b_{0,2} = 0\} \cap \{w =0\}, \nonumber\\
&& C_{ \mathbf 5^{(2)}} = \{c_{2,1}= 0\} \cap \{w =0\},\nonumber \\
&& C_{ \mathbf 5^{(3)}} = \{ c_1= 0\} \cap \{w =0\}, \\
&& C_{ \mathbf 5^{(4)}} = \{b_1 b_2 -  d_1 c_1=0\} \cap \{w =0\}, \nonumber \\
&& C_{ \mathbf 5^{(5)}} = \{ d_{2,2} b_1^{2} +  d_1( b_{0,2} d_1 -  d_{0,2} b_1)=0\} \cap \{w =0\}.\nonumber
\eea

From the top we read off  the resolution of the $SU(5)$ singularities. The hypersurface of the resolved 4-fold $\hat Y_4$ is described by the proper transform of \eqref{hyperSU51},  %
\begin{equation}
\begin{split}
0 = &  b_{0,2} e_{0}^2 e_{1} e_{4}  s_0^2 \T v^2  \T u +c_{2,1} e_{0} e_{1} e_{2}   s_0 \T w \T v^2  \\
&+ d_{0,2} e_{0}^2 e_{1} e_{3} e_{4}^{2}  \T v  s_0^2 s_1  \T u^{2} +  
   b_1  s_0 s_1 \T w \T v \T u +  \\
   &  +c_1 e_{1}e_{2}^2 e_{3}  \T w^{2} \T v s_1 +  d_{2,2} e_{0}^2 e_{1} e_{3}^2 e_{4}^{3}  s_0^2  s_1^2 \T u^{3} + \\
 & + d_1 e_{3} e_{4}  s_0 s_1^2 \T w \T u^{2} + b_2 e_{1} e_{2}^2 e_{3}^2 e_{4}  s_1^{2} \T w^2 \T u\,.  
\end{split}
\end{equation}
The resolution divisors $e_i=0$, $i=1, \ldots, 4$ are $\mathbb P^1$-fibrations over the GUT divisor and combine with $e_0$, the proper transform of $w=0$, into the affine Dynkin diagram of $SU(5)$.
As in the Weierstra\ss{} model, there are several possibilities to resolve the singularities.
For definiteness we choose a resolution whose 
SR-ideal includes, in addition to \eqref{SR1}, the elements
\begin{equation}
 \begin{split}
\{&  \T w  \, e_{0}, \T w \, e_{1}, \T w \, e_{3}, \T w \, e_{4}, s_1 \, e_{0},  s_1 \, e_{1}, s_1 \, e_{4}, s_0 \, e_{1},    s_0 \, e_{2}, \\
&s_0 \, e_{3},  s_0 \, e_{4},\T u \, e_{2}, e_{0} \, e_{2}, e_{2} \, e_{4},  
  \T v  \, e_{3}, \T v \, e_{4}, \T u \, e_{1},e_{0} \, e_{3}\}.  
 \end{split}
\end{equation}
Indeed it is now possible, with the techniques presented in \cite{Krause:2011xj}---see \cite{Esole:2011sm} for different approaches---to analyse the fibre splitting over the matter curves. This confirms the appearance of the affine Dynkin diagrams of $SO(10)$ and $SU(6)$ respectively. An explicit construction of the weight vectors associated with the respective representations proves  the appearance of the ${\mathbf{10}}$- and ${\mathbf 5}$-representations of $SU(5)$ at these loci.

Further, the generators $\tw_i \in H^{1,1}(\hat Y_4)$ must be modified such that the roots of $SU(5)$ are uncharged under $U(1)_i$. This amounts to the condition
\bea
\int_{\hat Y_4} \tw_i \wedge E_k \wedge \omega_b =0
\eea
in addition to transversality. Here $E_k$ are the 2-forms dual to the resolution divisors $\{e_k=0\}$. As in \cite{Krause:2011xj} we ensure this by adding suitable combinations of $E_i$. 
As a result of the intersection numbers
\begin{equation}
\int_{\hat Y_{4}} S_i \wedge E_{k} \wedge \omega_b { } = \delta_{kA} \int_{B_3}W \wedge \omega_b \\
\end{equation}
with $A=(0,3,4)$ for $i=(0,1,2)$ and $W$ the class associated with the GUT divisor on $B_3$, the solution is
\begin{equation}
\begin{split}
\T{w}_{1} &= 5( S_1 -  S_{0} - \mathcal{\bar K}) + m^{i}E_{i}\,,\\
\T{w}_{2} &= 5(\T S_2 - S_{0} - \mathcal{\bar K} - [ c_{1}] )+ l^{i}E_{i} \,,
\end{split}
\end{equation}
with $m^{i}=(2,4,6,3)^T$ and $l^{i}=(1,2,3,4)^T$. The overall normalisation is chosen to ensure integer charges.
The result of the computation of the $U(1)_i$ charges of the $SU(5)$ representations, as described in \cite{Krause:2011xj,Mayrhofer:2012zy}, is \cite{new}
\begin{equation}
\begin{split}
& C_{\mathbf{10}}:  {\mathbf{10}}_{-1,2} \\
 C_{\mathbf 5^{(1)}}:   {\mathbf 5}_{-3,1}  \quad & C_{\mathbf 5^{(2)}}:     {\mathbf 5}_{2,-4},  \quad  C_{\mathbf 5^{(3)}}:  {\mathbf 5}_{2,6},     \\
 C_{\mathbf 5^{(4)}}: &   {\mathbf 5}_{-3,-4}, \quad   C_{\mathbf 5^{(5)}}:  {\mathbf 5}_{2,1}.
\end{split}
\end{equation}
The complete charged massless matter spectrum consists of these states together with the $SU(5)$ singlets \eqref{Singlet-charges}. It is interesting to note that these charges  can be accommodated in a local 2-2-1 split spectral cover model as constructed in \cite{Dudas:2010zb,Dolan:2011iu}. The same applies to top 7 in appendix ~\ref{app-SU5tops1}.

At the intersection of the matter curves the fibre structure changes due to the split of some of the $\mathbb P^1$s and Yukawa interactions between the matter states 
are localised. 
The analysis of the $SU(5)$ charged interactions results in the following interactions:
\begin{center}
\begin{tabular}{c|c}
Point & Yukawa coupling \\
\hline
$\{w= b_1= c_{2,1}=0\}$ & ${\mathbf{10}}_{-1,2}  {\mathbf{10}}_{-1,2} {\mathbf 5}_{2,-4}$  \\
$\{ w= b_1= b_{0,2}=0\}$ & ${\mathbf{10}}_{-1,2} {\bar{\mathbf 5}}_{-2,-1} {\bar{\mathbf 5}}_{3,-1}$  \\
$\{ w= b_1= c_1=0\}$ & ${\mathbf{10}}_{-1,2} { \bar{\mathbf 5}}_{3,4} {\bar{\mathbf 5}}_{-2,-6}$ \\
$\{w= b_1= d_1=0\}$ & non-flat fibre
\end{tabular}

\end{center}

From the perspective of an $SU(5)$ GUT model, the structure of Yukawa couplings implies  ${\mathbf 5}_{H^u}= \mathbf{5}_{2,-4}$, while 
both $({\bar{\mathbf 5}}_m, {\bar{\mathbf 5}}_{H^d}) = (   \mathbf{\bar 5}_{-2,-1}, \mathbf{\bar 5}_{3,-1} )$ or $({\bar{\mathbf 5}}_m, {\bar{\mathbf 5}}_{H^d}) =  (\mathbf{\overline{5}}_{3,4},\mathbf{\overline{5}}_{-2,-6})$
(and the other way round in each case) are possible.
Extra Yukawa couplings exist between the charged singlets and the ${\mathbf 5}$-matter curves at the intersection points of the singlet curves with the GUT divisor $w=0$.
The pattern we find is in perfect agreement with the $U(1)$ charge assignments and will be presented in detail in \cite{new}.

Note that the fibre over the  points $\{w= b_1= d_1=0\}$ becomes 2-dimensional. If these points are present, the fibration is non-flat. 
To be on the safe side we can always restrict ourselves to fibrations over base spaces $B_3$ where the set $\{w= b_1= d_1=0\}$ is empty as a consequence of the intersection of the associated divisor classes. It is understood that the base $B_3$ has this property. This restriction is to be interpreted as a constraint on the base space to give rise to a well-defined F-theory compactification with the desired $SU(5) \times U(1) \times U(1)$ structure. The appearance of such a constraint is of course by no means unexpected and simply reflects the well-known phenomenon that  given a specific brane configuration not every compactification space is automatically compatible with it.

\subsection*{\texorpdfstring{$SU(5) \times U(1)$}{SU(5)xU(1)} via recombination}

The $U(1) \times U(1)$-fibrations presented in this article  lend themselves to studying brane recombination processes that Higgs the 
$U(1) \times U(1)$ to some linear combination. This way the $SU(5) \times U(1) \times U(1)$ models presented in this article %
give rise to a  large class of $SU(5) \times U(1)$ models.

Independently of the non-abelian gauge symmetry, we can 
consider e.g.\ the breaking of $U(1) \times U(1)$ to the sum of both abelian factors by giving a VEV to the singlet $\mathbf 1_{-5,5}$ and its conjugate localised along the curve $C_{\mathbf 1^{(1)}}$ in a D-flat manner. In brane language this corresponds to a recombination process that renders the fibration more generic. The analogous process interpolating from the models of \cite{Grimm:2010ez} with one $U(1)$ group to  fibrations  without abelian gauge symmetry is well understood \cite{Grimm:2010ez,Braun:2011zm,Krause:2012yh}. 
As it turns out, in the present situation a non-zero VEV for $\mathbf 1_{-5,5}$ plus conjugate corresponds to ``switching on'' the monomial $\T v^3$ in the cubic %
\eqref{eq:hyper1}. This leads to a ${\rm Bl}^1 \mathbb P^2[3]$-fibration.

Apart from breaking the abelian gauge group corresponding to the difference of the two $U(1)$ factors  this leads to a recombination of the matter curves intersecting $C_{\mathbf 1^{(1)}}$. This follows already from field theory in view of the presence of a corresponding Yukawa coupling involving $\mathbf 1_{-5,5}$. 
Therefore the two singlet curves $C_{\mathbf 1^{(4)}}$ and  $C_{\mathbf 1^{(5)}}$ combine into a single one along which one type of singlets of combined diagonal $U(1)$ charge $10$ lives. 
 
In the presence of $SU(5)$ singularities the recombination also affects the $SU(5)$ charge matter sector.
The  curve $C_{\mathbf 1^{(1)}}$ intersects the ${\mathbf 5}$-matter curves $C_{{\mathbf 5}^{(1)}}$ and $C_{{\mathbf 5}^{(2)}}$ in the point $w=b_{0,2} = c_{2,1}$, where the Yukawa coupling $\mathbf 1_{-5,5}   {\bar{\mathbf 5}}_{3,-1} {\mathbf 5}_{2,-4}$ is localised. A VEV for $\mathbf 1^{(1)}$ thus leads to the recombination of $C_{{\mathbf 5}^{(1)}}$ and $C_{{\mathbf 5}^{(2)}}$ into a single curve along which a ${\mathbf 5}$-state of diagonal charge $-2$ localises. 
All the remaining $SU(5)$ curves are unaffected, except that the states are charged only under the remaining diagonal $U(1)$. This structure of $SU(5)$ matter curves and their charges agrees with the corresponding data of the model in \cite{Braun:2013yti}. 
This is not unexpected since the top appearing in~\cite{Braun:2013yti} agrees with the top in the main part of this article, apart from the additional monomial $\alpha\,\tv^3$ which is responsible for the brane recombination. There are seven tops for such a ${\rm Bl}^1 \mathbb P^2[3]$-fibration, see~\cite{new}. Furthermore, note that the additional section of the  ${\rm Bl}^1 \mathbb P^2[3]$-fibration is not realised torically unlike the ${\rm Bl}^1 \mathbb P_{[1,1,2]}[4]$ case~\cite{Kreuzer:1997zg}.
More details will be presented in \cite{new}.

\section{Outlook}

Based on the polygon in FIG.~\ref{fig:polygon5} we have analysed the $8$ additional tops compatible with $SU(5)$ symmetry. 
The first six of these lead to a structure of one ${\mathbf{10}}$- and five ${\mathbf 5}$-matter curves and corresponding Yukawa points similar to the pattern presented in the previous section, albeit each with a different $U(1) \times U(1)$ charge assignment. We collect the main results of these models in appendix \ref{app-SU5tops1}.

In \cite{Mayrhofer:2012zy} we have described, inspired by \cite{Morrison:2012ei}, $SU(5) \times U(1)$-fibrations based on the elliptic fibre $\mathbb P_{[1,1,2]}[4]$.
This geometry likewise falls under the class of fibrations considered in \cite{Bouchard:2003bu}. In appendix \ref{AppB} we present the analogous data for $SU(5)$ tops associated with this geometry.

In \cite{Mayrhofer:2012zy} we were most interested, for the phenomenological reasons described therein, in fibrations with several ${\mathbf{10}}$-curves, for which we have provided also an alternative description based on a factorised Tate model. As it turned out, the appearance of more than one ${\mathbf{10}}$-curve requires a deviation from purely toric methods as these assume the base sections to be generic apart from factoring out overall powers of the GUT divisor. We found \cite{Mayrhofer:2012zy} that an $SU(4)$ model with subsequent deformations yields models of the desired type. A similar analysis for the ${\rm Bl}^2 \mathbb P^2[3]$-fibrations of this paper will be presented in \cite{new}.

\subsection*{Acknowledgements}
\vspace*{-0.4cm}We thank Thomas Grimm, Christian Pehle, Sakura Sch\"afer-Nameki and especially Harald Skarke for helpful discussions. The research of EP is supported by a Marie Curie Intra European Fellowship within the 7th European Community Framework Programme. The work of CM and TW was supported in part by the DFG under Transregio TR33 ``The Dark Universe''.

\appendix

\section{ \texorpdfstring{$SU(5)\times U(1)\times U(1)$}{SU(5)xU(1)xU(1)} fibrations} \label{app-SU5tops1}

In this appendix we briefly summarise the defining data and the resulting spectra for the remaining $SU(5)$ tops for the ${\rm Bl}^2 \mathbb P^2[3]$-fibration. 
We present here the proper transformed hypersurfaces of the resolved 4-fold $\hat Y_4$ describing the $SU(5) \times U(1) \times U(1)$ F-theory model as well as 
the $SU(5)$ matter spectrum and the Yukawa points. We also display where the fibration becomes non-flat in codimension-three. It is assumed that the base space $B_3$ chosen for the fibration does not allow these intersection loci such that the fibration remains flat everywhere. More details and derivations will appear in \cite{new}.
Note that in addition to the six classes of models presented in the sequel there are two more possible tops, which however exhibit unconventional enhancement loci %
so that we do not display them here.

\subsection*{Top 2}

\noindent\emph{Proper transform:}
\begin{equation*}
 \begin{split}
0 =&    b_{0,3} e_{0}^{3}e_{1}^{2}e_{2}e_{4}^{2}\tu \tv^{2} s_{0}^{2} + d_{0,2}e_{0}^{2} e_{1}^{2}e_{2}e_{4}\tu^{2}\tv s_{0}^{2}s_{1} +  \\
& d_{2,1}e_{0}e_{1}^{2}e_{2}\tu^{3}s_{0}^{2}s_{1}^{2} + c_{2,1}e_{0}e_{4}\tv^{2} \tw s_{0}   + b_{1}\tu \tv \tw s_{0} s_{1} +   \\
&  d_{1}e_{1}e_{2}e_{3}\tu^{2}\tw s_{0}s_{1}^{2} + c_{1}e_{2}e_{3}^{2}e_{4}\tv \tw^{2}s_{1} +\\
&b_{2}e_{1}e_{2}^{2}e_{3}^{3}e_{4} \tu \tw^{2} s_{1}^{2}   
 \end{split}
\end{equation*}
\noindent\emph{Matter curves and Yukawa points:}
\begin{center}
\begin{tabular}{c| c}
Curve on $\{w=0\}$ & Matter representation\\
\hline
$\{b_{1}=0\}$ & $\mathbf{10_{-1,-2}}$ + $\mathbf{\overline{10}_{1,2}}$\\
$\{c_{2,1}=0\}$ & $\mathbf{5_{-3,4}}$ + $\mathbf{\overline{5}_{3,-4}}$\\
$\{c_{1}=0\}$ & $\mathbf{5_{-3,-6}}$ + $\mathbf{\overline{5}_{3,6}}$\\
$\{d_{2,1}=0\}$ & $\mathbf{5_{-3,-1}}$ + $\mathbf{\overline{5}_{3,1}}$\\
$\{b_{1}b_{2} - c_{1}d_{1}=0\}$ & $\mathbf{5_{2,4}}$ + $\mathbf{\overline{5}_{-2,-4}}$\\
\parbox[c][10mm][c]{40mm}{\center  \vspace*{-4mm}$\{b_{0,3}b_{1}^{2} + c_{2,1}$\\$\times(c_{2,1}d_{2,1} - b_{1}d_{0,2})=0\}$} & $\mathbf{5_{2,-1}}$ + $\mathbf{\overline{5}_{-2,1}}$\\
\end{tabular}
\end{center}
\begin{center}
\begin{tabular}{c|c}
Point on $\{w=0\}$  & Yukawa coupling \\
\hline
$\{b_{1}=c_{1}=0\}$ & $\mathbf{\overline{10}_{1,2}} \mathbf{5_{2,4}} \mathbf{5_{-3,-6}}$  \\
$\{b_{1}=d_{1}=0\}$ & $\mathbf{10_{-1,-2}}\mathbf{10_{-1,-2}}\mathbf{5_{2,4}}$ \\
$\{b_{1}=d_{2,1}=0\}$ & $\mathbf{\overline{10}_{1,2}}\mathbf{5_{-3,-1}}\mathbf{5_{2,-1}}$  \\
$\{b_{1}=c_{2,1}=0\}$ & non-flat fibre \\
\end{tabular}
\end{center}

\subsection*{Top 3}
\noindent\emph{Proper transform:}
\begin{equation*}
 \begin{split}
0 &= b_{0,2} e_{0}^{2}e_{1}e_{4}\tu \tv^{2} s_{0}^{2} + d_{0,1}e_{0}e_{1}e_{2}\tu^{2}\tv s_{0}^{2}s_{1} + 
  c_{1}e_{3}e_{4}\tv \tw^{2}s_{1}   \nonumber   \\
& c_{2,2}e_{0}^{2}e_{1}e_{3}e_{4}^{2}\tv^{2} \tw s_{0}  + b_{1}\tu \tv \tw s_{0} s_{1} + d_{1}e_{1}e_{2}^{2}e_{3}\tu^{2}\tw s_{0}s_{1}^{2} + \nonumber \\
&   d_{2,1}e_{0}e_{1}^{2}e_{2}^{3}e_{3}\tu^{3}s_{0}^{2}s_{1}^{2} +b_{2}e_{1}e_{2}^{2}e_{3}^{2}e_{4} \tu \tw^{2} s_{1}^{2} \nonumber   
 \end{split}
\end{equation*}
\noindent\emph{Matter curves and Yukawa points:}
\begin{center}
\begin{tabular}{c|c}
Curve on $\{w=0\}$ & Matter representation\\
\hline
$\{b_{1}=0\}$ & $\mathbf{10_{1,1}}$ + $\mathbf{\overline{10}_{-1,-1}}$\\
$\{b_{0,2}=0\}$ & $\mathbf{5_{3,-2}}$ + $\mathbf{\overline{5}_{-3,2}}$\\
$\{c_{1}=0\}$ & $\mathbf{5_{-2,-7}}$ + $\mathbf{\overline{5}_{2,7}}$\\
$\{b_{0,2}c_{1} - b_{1}c_{2,2}=0\}$ & $\mathbf{5_{-2,3}}$ + $\mathbf{\overline{5}_{2,-3}}$\\
$\{c_{1}d_{1} - b_{1}b_{2}=0\}$ & $\mathbf{5_{3,3}}$ + $\mathbf{\overline{5}_{-3,-3}}$\\
$\{b_{1}d_{2,1} - d_{0,1}d_{1}=0\}$ & $\mathbf{5_{-2,-2}}$ + $\mathbf{\overline{5}_{2,2}}$\\
\end{tabular}
\end{center}
\begin{center}
\begin{tabular}{c|c}
Point on  $\{w=0\}$ & Yukawa coupling \\
\hline
$\{b_{1}=b_{0,2}=0\}$ & $\mathbf{\overline{10}_{-1,-1}} \mathbf{5_{3,-2}} \mathbf{5_{-2,3}}$    \\
$\{b_{1}=d_{0,1}=0\}$ & $\mathbf{\overline{10}_{-1,-1}}\mathbf{\overline{10}_{-1,-1}}\mathbf{\overline{5}_{2,2}}$   \\
$\{b_{1}=d_{1}=0\}$ & $\mathbf{\overline{10}_{-1,-1}} \mathbf{5_{3,3}} \mathbf{5_{-2,-2}}$   \\
$\{b_{1}=c_{1}=0\}$ & non-flat fibre  \\
\end{tabular}
\end{center}

\subsection*{Top 4}

\noindent\emph{Proper transform:}
\begin{equation*}
 \begin{split}
0 &= b_{0,3} e_{0}^{3}e_{1}^{2}e_{2}e_{4}\tu \tv^{2} s_{0}^{2} + d_{0,1}e_{0}e_{1}e_{2}\tu^{2}\tv s_{0}^{2}s_{1} + b_{1}\tu \tv \tw s_{0} s_{1}   \nonumber \\
& c_{2,2}e_{0}^{2}e_{1}e_{4}\tv^{2} \tw s_{0}+ d_{2}e_{1}e_{2}^{2}e_{3}\tu^{3}s_{0}^{2}s_{1}^{2}  +   \nonumber \\
&   d_{1}e_{1}e_{2}^{2}e_{3}^{2}e_{4}\tu^{2}\tw s_{0}s_{1}^{2} + c_{1}e_{3}e_{4}\tv \tw^{2}s_{1} +b_{2}e_{1}e_{2}^{2}e_{3}^{3}e_{4}^{2} \tu \tw^{2} s_{1}^{2} \nonumber  
 \end{split}
\end{equation*}
\noindent\emph{Matter curves and Yukawa points:}
\begin{center}
\begin{tabular}{c|c}
Curve on $\{w=0\}$ & Matter representation\\
\hline
$\{b_{1}=0\}$ & $\mathbf{10_{-1,1}}$ + $\mathbf{\overline{10}_{1,-1}}$\\
$\{c_{1}=0\}$ & $\mathbf{5_{-3,-7}}$ + $\mathbf{\overline{5}_{3,7}}$\\
$\{c_{2,2}=0\}$ & $\mathbf{5_{-3,3}}$ + $\mathbf{\overline{5}_{3,-3}}$\\
$\{d_{2}=0\}$ & $\mathbf{5_{-3,-2}}$ + $\mathbf{\overline{5}_{3,2}}$\\
$\{b_{1}^{2}b_{2} - b_{1}c_{1}d_{1} + c_{1}^{2}d_{2}=0\}$ & $\mathbf{5_{2,3}}$ + $\mathbf{\overline{5}_{-2,-3}}$\\
$\{b_{0,3}b_{1} - c_{2,2}d_{0,1}=0\}$ & $\mathbf{5_{2,-2}}$ + $\mathbf{\overline{5}_{-2,2}}$\\
\end{tabular}
\end{center}
\begin{center}
\begin{tabular}{c|c}
Point on $\{w=0\}$ & Yukawa coupling \\
\hline
$\{b_{1}=c_{2,2}=0\}$ & $\mathbf{10_{-1,1}} \mathbf{\overline{5}_{3,-3}} \mathbf{\overline{5}_{-2,2}}$ \\
$\{b_{1}=d_{0,1}=0\}$ & $\mathbf{\overline{10}_{1,-1}}\mathbf{\overline{10}_{1,-1}}\mathbf{\overline{5}_{-2,2}}$ \\
$\{b_{1}=d_{2}=0\}$ & $\mathbf{\overline{10}_{1,-1}} \mathbf{5_{-3,-2}} \mathbf{5_{2,3}}$ \\
$\{b_{1}=c_{1}=0\}$ & non-flat fibre \\
\end{tabular}
\end{center}

\subsection*{Top 5}

\noindent\emph{Proper transform:}
\begin{equation*}
 \begin{split}
&0 =  b_{0,2} e_{0}^{2}e_{1}e_{3}e_{4}^{2}\tu \tv^{2} s_{0}^{2} + d_{0,1}e_{0}e_{3}e_{4}^{2}\tu^{2}\tv s_{0}^{2}s_{1} +  b_{1}\tu \tv \tw s_{0} s_{1}   \nonumber \\
& + d_{2,1}e_{0}e_{1}e_{2}e_{3}^{2}e_{4}^{3}\tu^{3}s_{0}^{2}s_{1}^{2} + c_{2,1}e_{0}e_{1}\tv^{2} \tw s_{0} +  \nonumber \\
&  d_{1}e_{2}e_{3}e_{4}\tu^{2}\tw s_{0}s_{1}^{2} + c_{1,1}e_{0}e_{1}^{2}e_{2}^{2}e_{3}\tv \tw^{2}s_{1} +b_{2}e_{1}e_{2}^{2}e_{3} \tu \tw^{2} s_{1}^{2} \nonumber  
 \end{split}
\end{equation*}
\noindent\emph{Matter curves and Yukawa points:}
\begin{center}
\begin{tabular}{c|c}
Curve on $\{w=0\}$ & Matter representation\\
\hline
$\{b_{1}=0\}$ & $\mathbf{10_{-1,0}}$ + $\mathbf{\overline{10}_{1,0}}$\\
$\{b_{2}=0\}$ & $\mathbf{5_{-3,-1}}$ + $\mathbf{\overline{5}_{3,1}}$\\
$\{c_{2,1}=0\}$ & $\mathbf{5_{2,-1}}$ + $\mathbf{\overline{5}_{-2,1}}$\\
$\{b_{1}d_{2,1} -d_{0,1}d_{1}=0\}$ & $\mathbf{5_{2,0}}$ + $\mathbf{\overline{5}_{-2,0}}$\\
$\{b_{0,2}b_{1} - c_{2,1}d_{0,1}=0\}$ & $\mathbf{5_{-3,0}}$ + $\mathbf{\overline{5}_{3,0}}$\\
$\{b_{1}c_{1,1} - b_{2}c_{2,1}=0\}$ & $\mathbf{5_{2,1}}$ + $\mathbf{\overline{5}_{-2,-1}}$\\
\end{tabular}
\end{center}
\begin{center}
\begin{tabular}{c|c}
Point on $\{w=0\}$  & Yukawa coupling \\
\hline
$\{b_{1}=b_{2}=0\}$ & $\mathbf{\overline{10}_{1,0}} \mathbf{5_{2,1}} \mathbf{5_{-1,-3}}$  \\
$\{b_{1}=d_{1}=0\}$ & $\mathbf{\overline{10}_{1,0}}\mathbf{\overline{10}_{1,0}}\mathbf{\overline{5}_{-2,0}}$  \\
$\{b_{1}=d_{0}=0\}$ & $\mathbf{\overline{10}_{1,0}} \mathbf{5_{-3,0}}\mathbf{5_{2,0}}$  \\
$\{b_{1}=c_{2}=0\}$ & non-flat fibre  \\
\end{tabular}
\end{center}

\subsection*{Top 6}

\noindent\emph{Proper transform:}
\begin{equation*}
 \begin{split}
0 &=  b_{0,1} e_{0}e_{4}\tu \tv^{2} s_{0}^{2} + d_{0,1}e_{0}e_{2}e_{3}^{2}e_{4}^{2}\tu^{2}\tv s_{0}^{2}s_{1} + b_{1}\tu \tv \tw s_{0} s_{1} +   \nonumber \\
&  d_{2,1}e_{0}e_{2}^{2}e_{3}^{4}e_{4}^{3}\tu^{3}s_{0}^{2}s_{1}^{2} + c_{2,2}e_{0}^{2}e_{1}^{2}e_{2}e_{4}\tv^{2} \tw s_{0} + \nonumber \\
&  d_{1}e_{2}e_{3}^{2}e_{4}\tu^{2}\tw s_{0}s_{1}^{2} + c_{1,1}e_{0}e_{1}^{2}e_{2}\tv \tw^{2}s_{1} +b_{2}e_{1}e_{2}e_{3} \tu \tw^{2} s_{1}^{2}   \nonumber   
 \end{split}
\end{equation*}
\noindent\emph{Matter curves and Yukawa points:}
\begin{center}
\begin{tabular}{c|c}
Curve on $\{w=0\}$ & Matter representation\\
\hline
$\{b_{1}=0\}$ & $\mathbf{10_{2,2}}$ + $\mathbf{\overline{10}_{-2,-2}}$\\
$\{b_{0,1}=0\}$ & $\mathbf{5_{-4,1}}$ + $\mathbf{\overline{5}_{4,-1}}$\\
$\{b_{2}=0\}$ & $\mathbf{5_{-4,-4}}$ + $\mathbf{\overline{5}_{4,4}}$\\
$\{c_{1,1}=0\}$ & $\mathbf{5_{1,6}}$ + $\mathbf{\overline{5}_{-1,-6}}$\\
$\{b_{0,1}c_{1,1} - b_{1}c_{2,2}=0\}$ & $\mathbf{5_{1,-4}}$ + $\mathbf{\overline{5}_{-1,4}}$\\
$\{b_{0,1}d_{1}^{2} - b_{1}d_{0,1}d_{1} + b_{1}^{2}d_{2,1}=0\}$ & $\mathbf{5_{1,1}}$ + $\mathbf{\overline{5}_{-1,-1}}$\\
\end{tabular}
\end{center}
\begin{center}
\begin{tabular}{c|c}
Point on  $\{w=0\}$ & Yukawa coupling\\
\hline
$\{b_{1}=b_{2}=0\}$ & $\mathbf{10_{2,2}} \mathbf{10_{2,2}}\mathbf{5_{-4,-4}}$ \\
$\{b_{1}=c_{1,1}=0\}$ & $\mathbf{10_{2,2}} \mathbf{\overline{5}_{-1,4}}\mathbf{\overline{5}_{-1,-6}}$ \\
$\{b_{1}=b_{0,1}=0\}$ & non-flat fibre \\
\end{tabular}
\end{center}

\subsection*{Top 7}

\noindent\emph{Proper transform:}
\begin{equation}\nonumber 
 \begin{split}
&0 = b_{0,2} e_{0}^{2}e_{1}e_{3}e_{4}^{2}\tu \tv^{2} s_{0}^{2} + d_{0}e_{3}e_{4}\tu^{2}\tv s_{0}^{2}s_{1} +  b_{1}\tu \tv \tw s_{0} s_{1} +  \\
& d_{2}e_{1}e_{2}^{2}e_{3}^{3}e_{4}^{2}\tu^{3}s_{0}^{2}s_{1}^{2} + c_{2,2}e_{0}^{2}e_{1}e_{4}\tv^{2} \tw s_{0} \nonumber  \\ 
& +  d_{1}e_{1}e_{2}^{2}e_{3}^{2}e_{4}\tu^{2}\tw s_{0}s_{1}^{2} + c_{1,1}e_{0}e_{1}e_{2}\tv \tw^{2}s_{1} +b_{2}e_{1}e_{2}^{2}e_{3} \tu \tw^{2} s_{1}^{2} \nonumber  
 \end{split}
\end{equation}
\noindent\emph{Matter curves and Yukawa points:}
\begin{center}
\begin{tabular}{c|c}
Curve on $\{ w=0 \}$ & Matter representation\\
\hline
$\{b_{1}=0\}$ & $\mathbf{10_{-1,-3}}$ + $\mathbf{\overline{10}_{1,3}}$\\
$\{b_{2}=0\}$ & $\mathbf{5_{-3,-4}}$ + $\mathbf{\overline{5}_{3,4}}$\\
$\{c_{1,1}=0\}$ & $\mathbf{5_{2,6}}$ + $\mathbf{\overline{5}_{-2,-6}}$\\
$\{c_{2,2}=0\}$ & $\mathbf{5_{2,-4}}$ + $\mathbf{\overline{5}_{-2,4}}$\\
$\{b_{0,2}b_{1} - c_{2,2}d_{0}=0\}$ & $\mathbf{5_{-3,1}}$ + $\mathbf{\overline{5}_{3,-1}}$\\
$\{b_{2}d_{0}^{2} + b_{1}(b_{1}d_{2} - d_{0}d_{1})=0\}$ & $\mathbf{5_{2,1}}$ + $\mathbf{\overline{5}_{-2,-1}}$\\
\end{tabular}
\end{center}
\begin{center}
\begin{tabular}{c|c}
Point on $\{w=0\}$ & Yukawa coupling \\
\hline
$\{b_{1}=b_{2}=0\}$ & $\mathbf{10_{-1,-3}} \mathbf{\overline{5}_{-2,-1}} \mathbf{\overline{5}_{3,4}}$  \\
$\{b_{1}=c_{1,1}=0\}$ & $\mathbf{\overline{10}_{1,3}} \mathbf{\overline{10}_{1,3}}\mathbf{\overline{5}_{-2,-6}}$   \\
$\{b_{1}=c_{2,2}=0\}$ & $\mathbf{\overline{10}_{1,3}} \mathbf{5_{2,-4}}\mathbf{5_{-3,1}}$  \\
$\{b_{1}=d_{0}=0\}$ & non-flat fibre  \\
\end{tabular}
\end{center}

\section{\texorpdfstring{$SU(5) \times U(1)$}{SU(5)xU(1)} from \texorpdfstring{${\rm Bl}^1 \mathbb P_{[1,1,2]}[4]$}{Bl1 P[1,1,2][4]} } \label{AppB}

\begin{figure}[h]
\begin{flushright}
\def\svgwidth{0.4\textwidth}
 \executeiffilenewer{polygon6.svg}{polygon6.pdf}%
 {inkscape -z -D --file=polygon6.svg %
  --export-pdf=polygon6.pdf --export-latex}%
   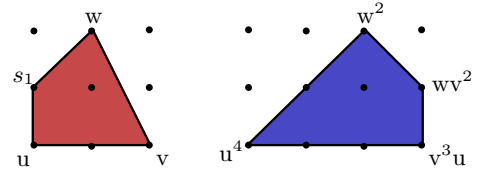%

      \caption{The toric polygon to $\textmd{Bl}^{1}\mathbb P_{[1,1,2]}$ and its dual. On the dual one we only indicated the monomials of the vertices and omitted powers of $s_1$. }\label{fig:polygon6}
\end{flushright}
\end{figure}

We now display the main results for the construction of $SU(5) \times U(1)$ F-theory compactifications  based on a  ${\rm Bl}^1 \mathbb P_{[1,1,2]}[4]$-fibration \cite{Morrison:2012ei}.
This is a direct continuation of our approach presented in \cite{Mayrhofer:2012zy}, to which we refer for details of the general setup.  While in \cite{Mayrhofer:2012zy} we were interested in $SU(4)$ singularities degenerating further to $SU(5)$, since these can accommodate multiple ${\bf 10}$-curves, here we summarise the result of a direct implementation of $SU(5)$ singularities via the tops construction.
For four tops 
we give the hypersurface of the resolved Calabi-Yau 4-fold $\hat Y_{4}$ as well as the $SU(5)$-matter spectrum and Yukawa points. The homogeneous coordinates of $\mathbb P_{[1,1,2]}$ are $[\tu : \tv : \tw]$ and $s_1$ denotes the rational section, cf.~FIG.~\ref{fig:polygon6}.
In addition, we list again the loci where the fibration becomes non-flat in codimension-three and point out that to obtain a suitable F-theory compactification one chooses the base $B_3$ such that these loci are absent. Note that there is a fifth possible top, which we have not included due to non-standard behaviour at the enhancement loci. We refer to \cite{new} for more details.

\subsection*{Top 1}

\noindent\emph{Proper transform:}
\begin{align*}
\begin{split}
0= &{ } \tw^{2}s_{1}e_{3}e_{4} + b_{0,2} \tw \tu^{2} s_{1}^{2} e_{0}^{2}e_{1}e_{4} + b_{1}\tu \tv \tw s_{1} \\&+ b_{2}\tv^{2} \tw  e_{1}e_{2}^{2}e_{3}^{2}e_{4} - c_{0,5}\tu^{4}s_{1}^{3}e_{0}^{5}e_{1}^{3}e_{2}e_{4}^{2}\\& - c_{1,3} \tu^{3} \tv s_{1}^{2} e_{0}^{3} e_{1}^{2}e_{2}e_{4} - c_{2,1}\tu^{2} \tv^{2}s_{1}e_{0}e_{1}e_{2}\\& - c_{3}\tu \tv^{3}e_{1}e_{2}^{2}e_{3}
\end{split}
\end{align*}
\noindent\emph{Matter curves and Yukawa points:}
\begin{center}
\begin{tabular}{cc}
Curve on $\{w=0\}$ & Matter representation\\
\hline
$\{b_{1}=0\}$ & $\mathbf{10_{0}}$ + $\mathbf{\overline{10}_{0}}$\\
$\{c_{3}=0\}$ & $\mathbf{5_{-1}}$ + $\mathbf{\overline{5}_{1}}$\\
$\{b_{1}b_{2} + c_{3}=0\}$ & $\mathbf{5_{1}}$ + $\mathbf{\overline{5}_{-1}}$\\
$\{b_{1}^{2}c_{0,5} - b_{0,2}b_{1}c_{1,3} + b_{0,2}^{2}c_{2,1}=0\}$ & $\mathbf{5_{0}}$ + $\mathbf{\overline{5}_{0}}$\\
\end{tabular}
\end{center}
\begin{center}
\begin{tabular}{c|c}
Point on $\{w=0\}$ & Yukawa coupling \\
\hline
$\{b_{1}=c_{2}=0\}$ & $\mathbf{\overline{10}_{0}} \mathbf{\overline{10}_{0}}\mathbf{\overline{5}_{0}}$   \\
$\{b_{1}=c_{3}=0\}$ & $\mathbf{10_{0}} \mathbf{\overline{5}_{1}} \mathbf{\overline{5}_{-1}}$  \\
\end{tabular}
\end{center}

\subsection*{Top 2}

\noindent\emph{Proper transform:}
\begin{align*}
\begin{split}
0= &{ } \tw^{2}s_{1}e_{3}e_{4} + b_{0,2} \tw \tu^{2} s_{1}^{2} e_{0}^{2}e_{1}e_{3}e_{4}^{2} + b_{1}\tu \tv \tw s_{1}  \\&+ b_{2}\tv^{2} \tw  e_{1}e_{2}^{2}e_{3} - c_{0,4}\tu^{4}s_{1}^{3}e_{0}^{4}e_{1}^{2}e_{3}e_{4}^{3}\\& - c_{1,2} \tu^{3} \tv s_{1}^{2} e_{0}^{2}e_{1}e_{4} - c_{2,1}\tu^{2} \tv^{2}s_{1}e_{0}e_{1}e_{2}\\& - c_{3,1}\tu \tv^{3}e_{0}e_{1}^{2}e_{2}^{3}e_{3}
\end{split}
\end{align*}
\noindent\emph{Matter curves and Yukawa points:}
\begin{center}
\begin{tabular}{cc}
Curve on $\{w=0\}$ & Matter representation\\
\hline
$\{b_{1}=0\}$ & $\mathbf{10_{2}}$ + $\mathbf{\overline{10}_{-2}}$\\
$\{b_{2}=0\}$ & $\mathbf{5_{6}}$ + $\mathbf{\overline{5}_{-6}}$\\
$\{b_{1}c_{3,1} + b_{2}c_{2,1}=0\}$ & $\mathbf{5_{-4}}$ + $\mathbf{\overline{5}_{4}}$\\
$\{b_{1}^{2}c_{0,4} - b_{0,2}b_{1}c_{1,2} - c_{1,2}^{2}=0\}$ & $\mathbf{5_{1}}$ + $\mathbf{\overline{5}_{-1}}$\\
\end{tabular}
\end{center}
\begin{center}
\begin{tabular}{c|c}
Point on $\{w=0\}$ & Yukawa coupling \\
\hline
$\{b_{1}=b_{2}=0\}$ & $\mathbf{\overline{10}_{-2}} \mathbf{5_{6}}\mathbf{5_{-4}}$   \\
$\{b_{1}=c_{1,2}=0\}$ & $\mathbf{\overline{10}_{-2}} \mathbf{5_{1}} \mathbf{5_{1}}$  \\
$\{b_{1}=c_{2,1}=0\}$ & $\mathbf{\overline{10}_{-2}}\mathbf{\overline{10}_{-2}}  \mathbf{\overline{5}_{4}}$  \\
\end{tabular}
\end{center}

\subsection*{Top 3}

\noindent\emph{Proper transform:}
\begin{align*}
\begin{split}
0= &{ } \tw^{2}s_{1}e_{3}^{2}e_{4} + b_{0,1} \tw \tu^{2} s_{1}^{2}e_{0}e_{4} + b_{1}\tu \tv \tw s_{1} \\&+ b_{2}\tv^{2} \tw e_{1}e_{2}e_{3} - c_{0,4}\tu^{4}s_{1}^{3}e_{0}^{4}e_{1}^{2}e_{2}e_{4}^{3}\\& - c_{1,3} \tu^{3} \tv s_{1}^{2} e_{0}^{3}e_{1}^{2}e_{2}e_{4}^{2} - c_{2,2}\tu^{2} \tv^{2}s_{1}e_{0}^{2}e_{1}^{2}e_{2}e_{4}\\& - c_{3,1}\tu \tv^{3}e_{0}e_{1}^{2}e_{2}
\end{split}
\end{align*}

\vspace*{1cm}
\noindent\emph{Matter curves and Yukawa points:}
\begin{center}
\begin{tabular}{cc}
Curve on $\{w=0\}$ & Matter representation\\
\hline
$\{b_{1}=0\}$ & $\mathbf{10_{-3}}$ + $\mathbf{\overline{10}_{3}}$\\
$\{b_{2}=0\}$ & $\mathbf{5_{6}}$ + $\mathbf{\overline{5}_{-6}}$\\
$\{c_{3,1}=0\}$ & $\mathbf{5_{-4}}$ + $\mathbf{\overline{5}_{4}}$\\
\parbox[c][10mm][c]{40mm}{\center  \vspace*{-4mm}$\{b_{1}^{3}c_{0,4} - b_{0,1}b_{1}^{2}c_{1,3} $\\ $ + b_{0,1}^{2}b_{1}c_{2,2} - b_{0,1}^{3}c_{3,1}=0\}$} & $\mathbf{5_{1}}$ + $\mathbf{\overline{5}_{-1}}$\\
\end{tabular}
\end{center}
\begin{center}
\begin{tabular}{c|c}
Point on $\{w=0\}$ & Yukawa coupling \\
\hline
$\{b_{1}=b_{2}=0\}$ & $\mathbf{10_{-3}} \mathbf{10_{-3}}\mathbf{5_{6}}$   \\
$\{b_{1}=c_{3,1}=0\}$ & $\mathbf{10_{-3}} \mathbf{\overline{5}_{-1}} \mathbf{\overline{5}_{4}}$  \\
$\{b_{1}=b_{0,1}=0\}$ & non-flat fibre\\
\end{tabular}
\end{center}

\subsection*{Top 4}

\noindent\emph{Proper transform:}
\begin{align*}
\begin{split}
0= &{ } \tw^{2}s_{1}e_{2}e_{3}^{2}e_{4} + b_{0,1} \tw \tu^{2} s_{1}^{2}e_{0}e_{3}e_{4} + b_{1}\tu \tv \tw s_{1} \\&+ b_{2}\tv^{2} \tw e_{1}e_{2} - c_{0,3}\tu^{4}s_{1}^{3}e_{0}^{3}e_{1}e_{3}e_{4}^{2}\\& - c_{1,2} \tu^{3} \tv s_{1}^{2} e_{0}^{2}e_{1}e_{4} - c_{2,2}\tu^{2} \tv^{2}s_{1}e_{0}^{2}e_{1}^{2}e_{2}e_{4}\\& - c_{3,2}\tu \tv^{3}e_{0}^{2}e_{1}^{3}e_{2}^{2}e_{4}
\end{split}
\end{align*}
\noindent\emph{Matter curves and Yukawa points:}
\begin{center}
\begin{tabular}{cc}
Curve on $\{w=0\}$ & Matter representation\\
\hline
$\{b_{1}=0\}$ & $\mathbf{10_{-1}}$ + $\mathbf{\overline{10}_{1}}$\\
$\{b_{2}=0\}$ & $\mathbf{5_{7}}$ + $\mathbf{\overline{5}_{-7}}$\\
$\{b_{1}c_{0,3} - b_{0,1}c_{1,2}=0\}$ & $\mathbf{5_{2}}$ + $\mathbf{\overline{5}_{-2}}$\\
$\{b_{2}^{2}c_{1,2} - b_{1}b_{2}c_{2,2} + b_{1}^{2}c_{3,2}=0\}$ & $\mathbf{5_{-3}}$ + $\mathbf{\overline{5}_{3}}$\\
\end{tabular}
\end{center}
\begin{center}
\begin{tabular}{c|c}
Point on $\{w=0\}$ & Yukawa coupling \\
\hline
$\{b_{1}=b_{0,1}=0\}$ & $\mathbf{\overline{10}_{1}} \mathbf{\overline{10}_{1}}\mathbf{\overline{5}_{-2}}$   \\
$\{b_{1}=c_{1,2}=0\}$ & $\mathbf{\overline{10}_{1}} \mathbf{5_{-3}} \mathbf{5_{2}}$  \\
$\{b_{1}=b_{2}=0\}$ & non-flat fibre\\
\end{tabular}
\end{center}

\end{document}